**Critical damping in linear system with two degrees of freedom: surprises and pitfalls.**

Maxim D. Arnold[1], Oleg V. Gendelman[2,*] and Vadim Zharnitsky[3]

1 – Department of Mathematical Sciences, The University of Texas at Dallas, Richardson, USA, e-mail: maxim.arnold@utdallas.edu

2 – Faculty of Mechanical Engineering, Technion – Israel Institute of Technology, Haifa, Israel, e-mail: ovgend@technion.ac.il

3 – Department of Mathematics and Coordinated Sciences Laboratory, University of Illinois – Urbana-Champaign, USA, e-mail: vzh@illinois.edu

* - contacting author

**In this Brief Communication, we establish the notion of critical damping for generic two-degree-of-freedom system. For given set of masses and stiffnesses, the critical damping corresponds to the real eigenvalue with maximal multiplicity, equal to minus geometrical mean of the eigenfrequencies. This case corresponds to the fastest possible asymptotic decay rate for generic initial conditions. The damping matrix for the critical case is unique up to reflection of one modal coordinate and, generically, non-diagonal. Quite surprisingly, for large difference of the eigenfrequencies, it is also not positive definite – therefore, physical realization of the critical case will require active elements that provide negative effective damping.**

Classical model describing the motion of lumped masses coupled by linear elastic and linear damping elements is one of most popular benchmarks in numerous branches of mechanics, physics and engineering [1 - 3]. In canonical form, it is written as

$$\mathbf{M}\ddot{\mathbf{x}}+\mathbf{C}\dot{\mathbf{x}}+\mathbf{K}\mathbf{x}=0 \tag{1}$$

In purely mechanical setting, usually **M** is mass matrix, **C** is damping matrix, **K** is stiffness matrix and **x** is the displacement vector of the system. In other (e.g. electromechanical) settings the meaning of coefficients can be very different. In the case **C**=0 and positive definite **M** and **K**, system (1) reduces to a set of non-interacting oscillatory modes. If the damping matrix is diagonal in the modal coordinates [4], one deals with non-interacting decaying modes. For generic **C**, it is not the case, and one should deal with state-space representation and complex modes [5]. Such setting can possess many interesting properties absent in classical modal picture, and attracts substantial attention [6].

In single -degree-of-freedom (SDOF) linear oscillatory systems the notion of critical damping is a basic textbook material [1, 2]. In normalized form, the equation of motion is written as:

$$\ddot{q} + 2\xi\omega\dot{q} + \omega^2 q = 0 \quad (2)$$

Here $\omega > 0$ is a natural frequency, $\xi \geq 0$ is the damping factor. It is well known that the case $\xi = 1$ is critical - the eigenvalues of equation (2) in this case are repetitive, $\lambda_{1,2}^* = -\omega$, and for any other value of $\xi$ at least one eigenvalue will have bigger real part:

$$\max(\mathrm{Re}(\lambda_1), \mathrm{Re}(\lambda_2)) > -\omega \quad (3)$$

It means that if the damping is different from the critical value, for some initial conditions the excitation will asymptotically (at large time instances) decay slower than in the critical case.

We would like to explore the possibility of extending the notion of the critical damping for generic 2DOF damped system. To set the stage, we consider the generic 2DOF system of masses, springs and linear damping elements (Figure 1).

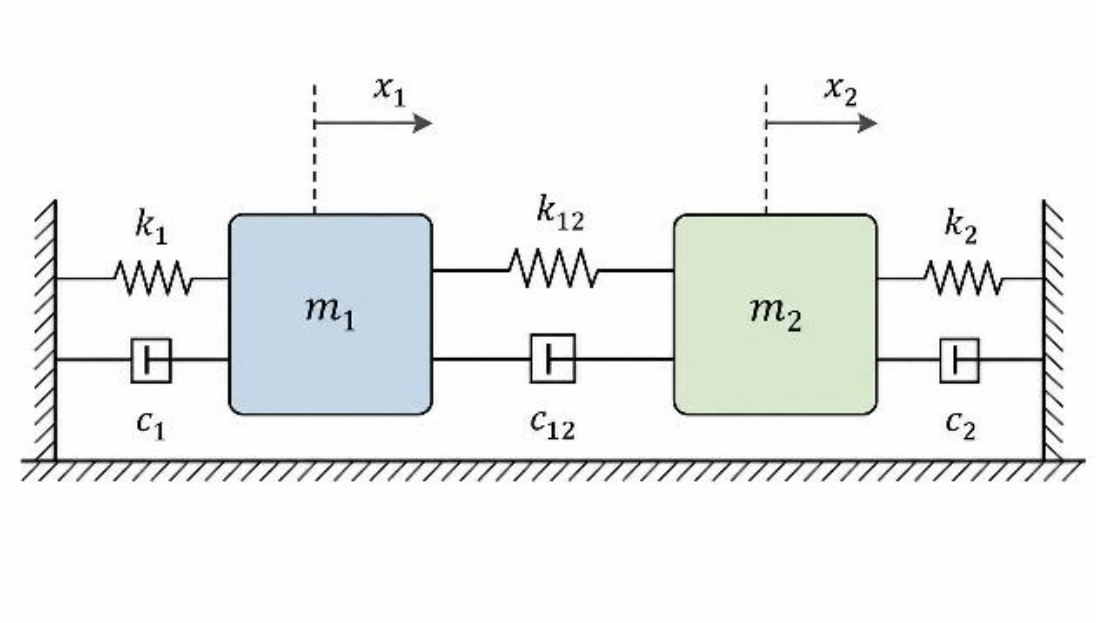

*Figure 1. Sketch of 2DOF coupled damped system.*

The notations are clear from the sketch. As it is well-known [4,5], by appropriate orthogonal coordinate transformation equations of motion of this model can be cast in the following generic form:

$$\ddot{\mathbf{q}} + 2\mathbf{\Gamma}\dot{\mathbf{q}} + \mathbf{\Omega}^2\mathbf{q} = 0, \ \mathbf{q} = \begin{pmatrix} q_1 \\ q_2 \end{pmatrix} \quad (4)$$

Here $\mathbf{\Omega} = \mathrm{diag}(\omega_1, \omega_2), \ \omega_{1,2} > 0$ is the frequency matrix, $\mathbf{\Gamma} = \begin{pmatrix} \gamma_{11} & \gamma_{12} \\ \gamma_{12} & \gamma_{22} \end{pmatrix}$ is a symmetric damping matrix. Then, for given frequency matrix, one can ask the following question: whether it is possible to find the entries in the damping matrix that will endow the eigenvalues of Eq. (3) with the property of criticality like the single-DOF case? Assume all eigenvalues of (4) have negative real parts. For generic initial conditions, the asymptotic decay rate of excitation is governed by the eigenvalue with ***maximal*** real part. The critical regime is defined as one where the ***maximal*** real part of four eigenvalues of the problem achieves ***minimal*** value possible for given pair $(\omega_1, \omega_2)$. Such property will provide the fastest possible asymptotic decay rate for given values of masses and stiffnesses.

Characteristic polynomial of System (4) is written as follows:

$$P(\lambda) = \lambda^4 + 2\lambda^3(\gamma_{11} + \gamma_{22}) + \lambda^2(\omega_1^2 + \omega_2^2 + 4(\gamma_{11}\gamma_{22} - \gamma_{12}^2)) + 2\lambda(\gamma_{11}\omega_2^2 + \gamma_{22}\omega_1^2) + \omega_1^2\omega_2^2 \quad (5)$$

According to Vieta theorem, the roots of $P(\lambda)$ obey the relationship

$$\lambda_1\lambda_2\lambda_3\lambda_4 = \omega_1^2\omega_2^2 \quad (6)$$

We assume that all eigenvalues have negative real parts. In this case, it is easy to see that the minimum of maximal real part of the set of eigenvalues is achieved if all eigenvalues are equal to negative geometric mean of the frequencies:

$$\min_{\Gamma}(\max_j(\mathrm{Re}(\lambda_j))) = -\sqrt{\omega_1\omega_2},\ \mathrm{j} = 1..4 \quad (7)$$

Indeed, if all eigenvalues are real and not equal, then, according to (5), at least one of them will be closer to zero than $\sqrt{\omega_1\omega_2}$. If, otherwise, some pair of eigenvalues is complex conjugate (say, $\lambda_{1,2} = \alpha \pm i\beta$), then it will contribute to (6) as $\lambda_1\lambda_2 = \alpha^2 + \beta^2$ and therefore at least one of remaining eigenvalues or both $\lambda_{1,2}$ will be closer to zero (will have larger real parts) than $\sqrt{\omega_1\omega_2}$. Thus, similarly to the single-DOF case, the condition of criticality is achieved for maximal multiplicity of the root of the characteristic polynomial.

This condition of the maximal multiplicity uniquely defines the characteristic polynomial:

$$\lambda_{1,..,4} = -\sqrt{\omega_1\omega_2};\ \ P(\lambda) = (\lambda + \sqrt{\omega_1\omega_2})^4 \quad (8)$$

Comparing term-wise with (5), one easily obtains the following solution for coefficients of the damping matrix *Γ*:

$$\gamma_{11} = \frac{2\omega_1^{3/2}\omega_2^{1/2}}{\omega_1 + \omega_2};\ \gamma_{22} = \frac{2\omega_1^{1/2}\omega_2^{3/2}}{\omega_1 + \omega_2};\ \gamma_{12} = \pm\frac{(\omega_2 - \omega_1)^2}{2(\omega_1 + \omega_2)} \quad (9)$$

Solution (9) is unique in a sense that the sign change in $\gamma_{12}$ is equivalent to simple sign reflection on one of modal coordinates $(q_1, q_2)$.

One can note, first, that critical damping matrix (9) is ***not diagonal***, besides simple degenerate case $\omega_1 = \omega_2$. In other terms, generically the critical damping is not proportional. It is not surprising – according to (3), for the proportional damping $\max(\mathrm{Re}(\lambda_j)) = -\min(\omega_1, \omega_2)$

. Thus, for generic initial conditions and non-degenerate system the proportional damping will indeed provide slower asymptotic decay rate.

To analyze the properties of the critical damping (9), it is instructive to look at dynamics of energy decay. For System (4), one naturally defines instantaneous energy:

$$E = \frac{1}{2}\left(\dot{q}_1^2 + \dot{q}_2^2 + \omega_1^2 q_1^2 + \omega_2^2 q_2^2\right) \tag{10}$$

From (4), one easily obtains:

$$\frac{dE}{dt} = -2\begin{pmatrix}\dot{q}_1 & \dot{q}_2\end{pmatrix}\mathbf{\Gamma}\begin{pmatrix}\dot{q}_1 \\ \dot{q}_2\end{pmatrix} = -2\left(\gamma_{11}\dot{q}_1^2 + \gamma_{22}\dot{q}_2^2 + 2\gamma_{12}\dot{q}_1\dot{q}_2\right) \tag{11}$$

Condition (8) guarantees that any excitation asymptotically decays to zero. However, from (11) one infers that the energy will decay monotonously for any excitation, only if the damping matrix is positive definite. From (9), it immediately follows that this condition is satisfied if the frequency ratio falls into the following interval:

$$3 - \sqrt{8} < \frac{\omega_1}{\omega_2} < 3 + \sqrt{8} \tag{12}$$

It means that outside this interval of frequency ratios the stiffness matrix for the case of critical damping will not be positive definite. Therefore, for some initial conditions, the energy will not decrease monotonously – one will observe temporary increases. Then, to realize such critically damped system physically, one must use some elements with negative damping coefficient – for instance, by means of active control. It is a striking difference with respect to the critical damping for singe degree of freedom.

To illustrate this point, we present in Figure 2 the time series for the instantaneous energy for two frequency ratios – inside and outside interval (12). Besides, the energy decay for the case of global critical damping is compared to the decay for proportional critical damping (see caption for the details).

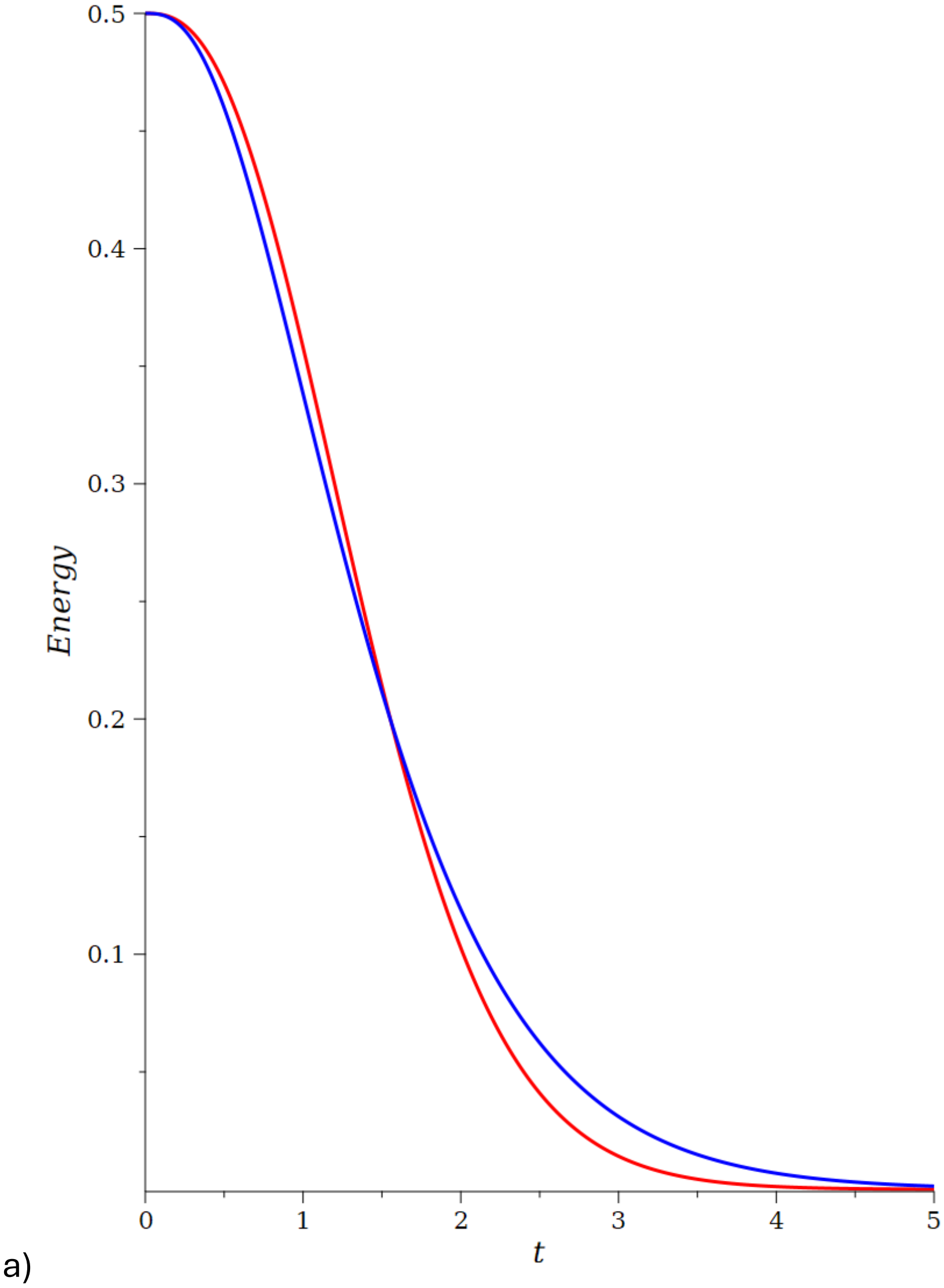

a)

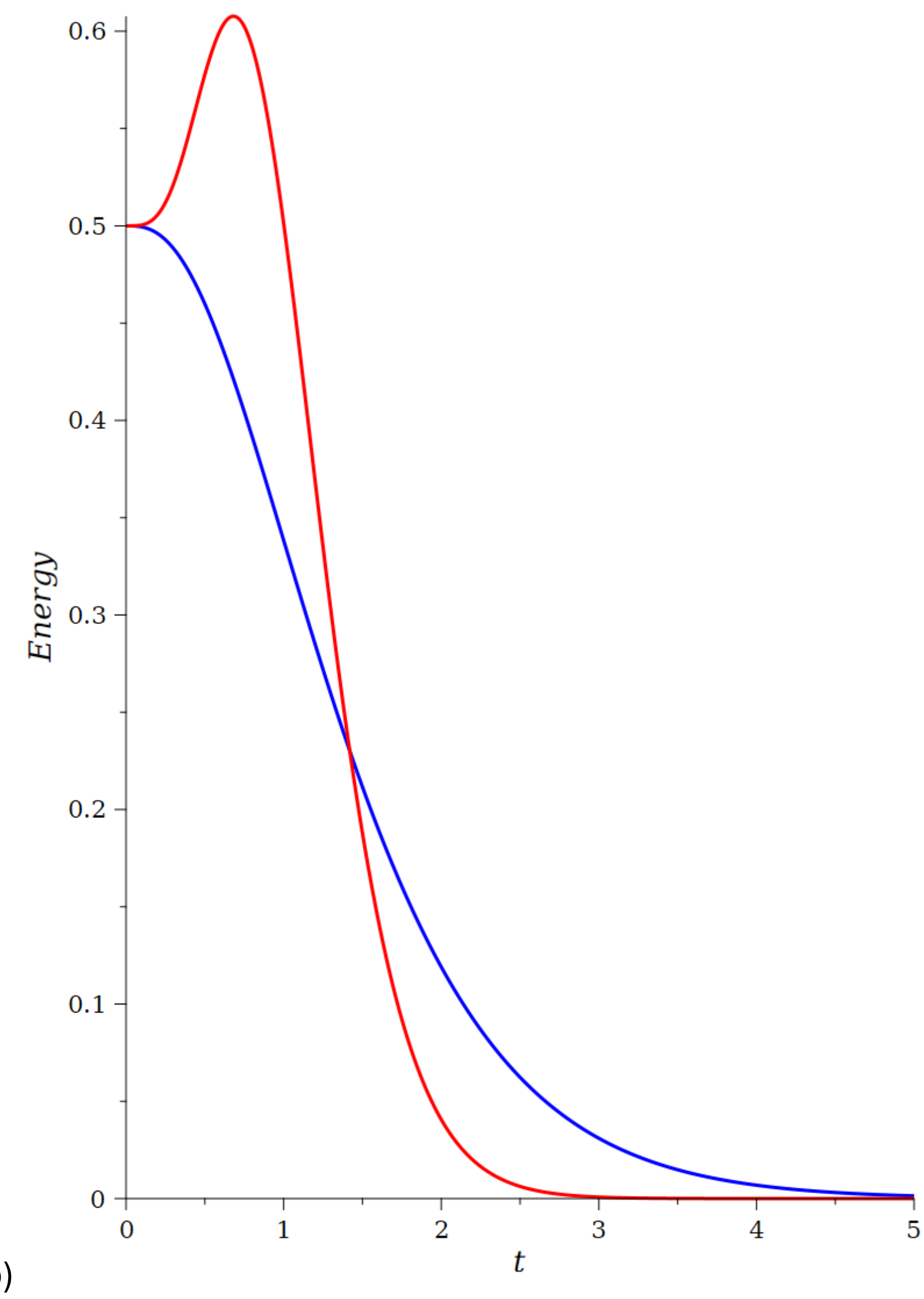

b)

*Figure 2. Instantaneous energy versus time for different damping matrices. Red – global critical damping (8-9), blue – proportional critical damping* ($\gamma_{11} = \omega_1, \gamma_{22} = \omega_2, \gamma_{12} = 0$)*. Initial conditions:* $q_1(0) = 1,\ \dot{q}_1(0) = q_2(0) = \dot{q}_2(0) = 0$*. a)* $\omega_1 = 1,\ \omega_2 = 4$*; b)* $\omega_1 = 1,\ \omega_2 = 10$

One indeed observes that the global critical damping provides faster asymptotic decay rate than the proportional critical damping. However, for large asymmetry of the eigenfrequencies the damping matrix (9) is not positive definite, and active elements with negative effective damping should be used.

To conclude, the results presented in this Brief Communication demonstrate that the notion of global critical damping can be extended for 2DOF system. This case corresponds to real eigenvalue with maximal multiplicity, equal to minus the geometric mean of the eigenfrequencies. The damping matrix that corresponds to this case can be established in unique way, up to reflection of one of the modal coordinates.

This global critical damping exhibits interesting, unexpected and maybe somewhat disappointing property – if the asymmetry of the eigenfrequencies is strong enough, the damping matrix ceases to be positive definite and physical realization of this scheme must include active elements with negative effective damping. To our opinion, this property is surprising but not discouraging – for some applications, the designer can decide to invest in appropriate control scheme to achieve the fastest possible decay rate for arbitrary initial conditions.

Natural question to address is the extension to arbitrary set with many degrees of freedom. It is easy to obtain system similar to (4) and to conclude that the global critical damping will correspond to the eigenvalue with maximal multiplicity, equal to minus the geometric mean of the eigenfrequencies – by consideration similar to the 2DOF case. However, the appropriate coefficients of the damping matrix cannot be found by simple term-wise inspection of the characteristic polynomial. This problem (including also existence/uniqueness question for the coefficients) requires more elaborate mathematical technique and will be analyzed elsewhere.

O.V.G. is grateful to Israel Science Foundation (grant 3085/25) for financial support.